\documentstyle[11pt]{article}
\begin{document}
\vspace{0.5cm} \centerline{\Large \bf Note on color neutral
solutions of the $K^0$ condensed } \centerline{\Large\bf
color-flavor locked phase}
 \vspace{1cm} \centerline{Xiao-Bing Zhang
} \vspace{0.1cm} \centerline{\small Department of Physics, Nankai
University, Tianjin 300071, China}  \vspace{8pt}

\vspace{0.5cm}

\begin{minipage}{13cm}
{\rm \noindent In the presence of nonzero strange quark mass $m_s$,
we investigate color neutrality in the $K^0$ condensed phase of
color-flavor locked quark matter. By treating the $m_s$ effects on
both kaon condensate and Fermi-surface phenomenon self-consistently,
we develop a new treatment to evaluate color neutral solutions
within the model-independent framework. It is pointed out that, in
the general sense, the expectation values of gluon fields obtained
from dynamics of Goldstone bosons solely are not identified with the
factual color chemical potentials. }

\vspace{0.5cm} {\bf PACS number(s): 25.75.Nq, 12.39.Fe, 12.38.-t}

\end{minipage}

\baselineskip 18pt

At very high densities, the lowest energy state of three-flavor,
three-color quark matter is widely expected to be the color-flavor
locked ( CFL ) phase, where the original color and flavor
$SU(3)_{color}\times SU(3)_{L} \times SU(3)_{R}$ symmetries of QCD
are broken down to the diagonal subgroup $SU(3)_{color+L+R}$
\cite{alf99,sch00,evans00}. In the ideal situation where flavor
asymmetry of quarks is ignored, the CFL matter should not carry any
color charges so that color neutrality is realized automatically. In
realistic situations, flavor asymmetry leads to some differences
among the Fermi momenta for nine ( flavor 3 $\times$ color 3) paired
quarks. In the CFL matter, nonzero color and electric charges may
appear and color/electric neutral problem needs to examined
seriously. In the presence of strange quark mass $m_s$, Alford and
Ragagopal have investigated color/electric neutrality by introducing
the chemical potentials associated with color charges \cite{alfj02}.
As far as color is concerned, they found that the nontrivial values
of color chemical potential ( the so-called color neutral solutions
) cancels the $m_s$-induced color charges and guarantees the
enforced neutrality of the bulk CFL matter. In the presence of
$m_s$, on the other hand, a less-symmetric CFL phase has been
predicted where the Goldstone bosons related to the
color-flavor-locked symmetry pattern become condensed \cite{sch,kr}.
For the CFL phase with maximal $K^0$ condensate, Kryjeski has
discussed its color neutrality by using a different method : the
color neutral solutions in the form of the expectation values of
gluon fields were solved from dynamics of Goldstone bosons
\cite{kry03}.

In the present note, we would like to reexamine color neutral
problem in the CFL matter with \emph{arbitrary} $K^0$ condensate (
hereafter CFL$K^0$ for short). The question needs to be answered,
whether the method used by Alford and Rajagopal in the conventional
CFL phase play its role in the CFL$K^0$ case. In other words, it is
whether the Kryjevski method used in the maximal $K^0$ condensed
phase be sufficient for color neutral solutions. We will point out
that, for a general CFL$K^0$ phase, the Kryjevski-type solutions do
not correspond to the Alford-Rajagopal-defined chemical potentials
strictly. Then, the factual color neutral solutions in $K^0$
condensed environment are obtained by treating the $m_s$ effects
self-consistently. These conclusions clarify the links between the
two kinds of color neutral solutions and correct the popular
suppositions on color neutral problem.

Before going to specifics, let 's briefly review the basic lines of
Refs. \cite{alfj02} and \cite{kry03}. Within the model-independent
framework, Alford and Ragagopal introduced the chemical potentials
coupled with diagonal generators of $SU(3)_{color}$ to examine the
CFL color neutrality in the presence of $m_s$
\cite{alfj02}.\footnotemark[1] \footnotetext[1] { As stressed in
Ref.\cite{rw01}, no electrons are required for the CFL matter
despite the unequal quark masses. Even in the presence of an
electric chemical potential, electric neutrality is easier realized
with respect to color neutral problem. For simplicity, we ignore the
electron chemical potential and only concern color neutrality all
through this work.  }  Assuming that the strange quark is far
heavier than the light quarks ( $m_s>>m_{u,d}$ ) and the quark
chemical potential is far larger than the quark masses ( $\mu>>m_s$
), the mismatch between the Fermi momenta for strange- and
light-flavor quarks is equal to ${m_s^2}/{2\mu}$ at the leading
order. For four kinds of quark pairs, the common Fermi momenta were
found to be \cite{alfj02}
\begin{eqnarray}
& & p_{F,(gs,bd)}^{com}=\mu -\frac{1}{4}\mu_3 -\frac{1}{6} \mu_8
-\frac{m_s^2}{4\mu}, \nonumber\\
& & p_{F,(rs,bu)}^{com}=\mu +\frac{1}{4}\mu_3 -\frac{1}{6} \mu_8
-\frac{m_s^2}{4\mu},\nonumber \\
& & p_{F,(rd,gu)}^{com}=\mu +\frac{1}{3} \mu_8
,\nonumber \\
& & p_{F,(ru,gd,bs)}^{com}=\mu -\frac{m_s^2}{6\mu},
 \label{pfcommon}
\end{eqnarray}
where $\mu_3$ and $\mu_8$ are defined to be coupled with the
generators $T_3=\lambda_3/2$ and $T_8=\lambda_8/\sqrt{3}$
respectively ( $\lambda_\alpha$ denotes the Gell-Mann matrices ).
Based on the Fermi surface phenomenon described by
Eq.(\ref{pfcommon}), they obtained the color neutral solutions
 \begin{eqnarray} \mu_3=0, \,\,\,\,\,\,
\mu_8=-\frac{m_s^2}{2\mu}, \label{ncfl}
\end{eqnarray}
from vanishing derivatives of the CFL free energy with respect to
$\mu_3$ and $\mu_8$. By inserting Eq.(\ref{ncfl}) into
(\ref{pfcommon}), it is found that the Fermi momenta for all pairs
are reduced to $p_{F}^{com}=\mu -\frac{m_s^2}{6\mu}$ which
corresponds to the Fermi surface phenomenon in the color-neutral CFL
phase.

Different from the Alford-Ragagopal method, the Kryjeski introduced
color static potentials are the expectation values of gluon fields,
namely the gluon condensates. In general, our discussion does not
limited in the case of the maximal $K^0$ condensate studied in Ref.
\cite{kry03}. For this purpose, it is practical to consider the
effective chemical potential associated with strangeness $\mu_S$
that equal to the $K^0$-mode chemical potential due to the chemical
equilibrium. At the leading order, $\mu_S$ might be given by
$m_s^2/2\mu$   ( if without the external strangeness chemical
potential \cite{zhang03} ). Once $\mu_{S}$ exceeds the kaon mass
$m_{K^0}$, $K^0$ condensation occurs and its condensate strength is
characterized by $\cos\theta=m_{K^0}^2/\mu_{S}^2$. Correspondingly,
the chiral field $\Sigma=\exp[\frac{i}{f_\pi}\lambda_\alpha
\pi_\alpha]$ ( $\pi_\alpha$ denotes the Goldstone octet and $f_\pi$
is the in-CFL-medium decay constant ) takes the form of \cite{kr}
\begin{eqnarray}
 \Sigma_{K^0}=\left(
\begin{array}{llcl}
1&0&0\\0&\cos\theta &i\sin\theta \\0&i\sin\theta&\cos\theta
\end{array}
\right)
 \label{sigmak}
\end{eqnarray}
in chiral effective Lagrangian accounting for the Goldstone bosons.
By extrapolating the original Kryjeski treatment, we find that the
temporal components of color-diagonal gluon fields have the nonzero
expectation values
\begin{eqnarray} & & -g A^0_3=
-\frac{m_s^2}{4\mu}+ \frac{m_s^2}{4\mu}\cos\theta, \nonumber
\\
& & -\frac{\sqrt{3}}{2}g A^0_8=
-\frac{m_s^2}{8\mu}-\frac{3m_s^2}{8\mu}\cos\theta,
 \label{A38}
\end{eqnarray}
where $g$ is the QCD coupling coefficient. The above result reflects
the anisotropic " vacuum " for Goldstone excitations, since it  is
obtained from the free energy of the Goldstone bosons solely.

As above mentioned, our concerned question is actually whether
Eq.(\ref{A38}) be adequate for the CFL$K^0$ color neutrality. Note
that a gluon field should attach to the quark loop, which accounts
for the color charge density at mean-field level. Therefore, it is
reasonable that the expectations values
 $g A^0_{3}$ and $g A^0_{8}$ behave like the color chemical potentials
$\mu_{3}$ and $\mu_{8}$ respectively. Superficially, one might
identify the two kinds of color static potentials with each other.
In the literature, the assumption like
\begin{eqnarray} \mu_3 = -g A^0_3, \,\,\,\,\,\,
\mu_8 = -\frac{\sqrt{3}}{2}g A^0_8, \label{link}
\end{eqnarray}
has been widely adopted, where the factor $\frac{\sqrt{3}}{2}$
arises to adapt the non-standard form of the color generator $T_8$
in Ref.\cite{alfj02}. In the maximal $K^0$ condensed case (
$\theta=\pi/2$ ), for instance, $-g A^0_3= -{m_s^2}/{4\mu}$ and
$-\sqrt{3}g A^0_8/2= -{m_s^2}/{8\mu}$  were supposed to be equal to
the color chemical potentials $\mu_{3}$ and $\mu_{8}$ respectively
\cite{forbes,zhang}. With the help of Eq.(\ref{link}),
the above question seems to have been well answered and the two
kinds of methods seems equivalent towards the CFL$K^0$ neutral
solutions. However, it is not the whole story yet. Eq.(\ref{link})
itself does not mean the one-to-one correspondence between the two
kinds of color neutral solutions ( although it might be available in
somewhat conditions ). In fact, Alford and Rajagopal evaluated color
neutrality from the Fermion free energy while Kryjevski did from the
free energy of Goldstone bosons. Since quarks and Goldstone bosons
belong to the distinct degrees of freedom, there is no a direct
reason to identify the two kinds of solutions with each other.
Secondly and perhaps more importantly, the Goldstone-mode
condensation makes sense in the vicinity of the Fermi surface of CFL
quark matter. This implies that the Goldstone-excitation vacuum
Eq.(\ref{A38}) becomes possible for a specific Fermi surface
phenomenon, i. e. a kind of Fermion accumulation made up of the
paired quarks. According to the Alford-Rajagopal method,
nevertheless, Fermi surface phenomena are relevant for color
neutrality also. Indeed, it is mismatches in the Fermi momentum
space to induce color chemical potentials essentially. In this case,
the factual neutral solutions should be determined by not only
Eq.(\ref{A38}) obtained from chiral Lagrangian solely but also the
Alford-Rajagopal-type solutions. Therefore, the correspondence which
is realized by Eq.(\ref{link}) is problematic, unless the Fermion
accumulation in $K^0$ condensed environment is proved irrelevant for
color neutrality.

Now we do not presume Eq.(\ref{link}) but keep it in mind that the
Kryjevski 's ( expectation values of ) gluon fields play the similar
roles as the color chemical potentials. As a starting point, we
suppose that $-g A^0_3$ and $-\frac{\sqrt{3}}{2}g A^0_8$ are only
the \emph{parts} of $\mu_3$ and $\mu_8$ respectively while the
remaining parts are given by the color static potentials $\mu'_3$
and $\mu'_8$ respectively. Explicitly, we define the latter as
\begin{eqnarray} \mu'_3=\mu_3 + g A^0_3,\,\,\,\,\,\,\,\,
\mu'_8=\mu_8+\frac{\sqrt{3}}{2}g A^0_8,\label{nu'}
\end{eqnarray}
which actually generalizes Eq.(\ref{link}). Note that the total free
energy for the CFL$K^0$ phase consists of the contributions from the
Fermion accumulation, the diquark condensates as well as the
Goldstone bosons. In Ref.\cite{alfj02}, the color superconducting
gaps were treated as parameters so that the second contribution (
the free energy from diquark condensates ) is not important. As for
the third contribution ( the free energy of Goldstone bosons ), it
has been used in yielding the expectation values of gluon fields. To
evaluate the factual color neutral solutions expressed by
$\mu_{3,8}$ eventually, we will solve the fictional chemical
potentials $\mu'_{3,8}$ from the first contribution ( the Fermion
free energy ) along the line of Ref.\cite{alfj02} while taking the
Kryjevski-type solutions Eq.(\ref{A38}) into account simultaneously.
As shall be seen below, $\mu'_{3,8}$ have usually the nonzero values
and their introduction is relevant for studies of the CFL$K^0$ color
neutrality.

As functions of $\mu'_{3,8}$ and $A^0_{3,8}$, the chemical
potentials for the color-flavor locked quarks read
\begin{eqnarray}
& &\mu_{bd}=[\mu-\frac{2}{3}(-\frac{\sqrt{3}}{2}g
A^0_8)]-\frac{2}{3}\mu'_8 , \nonumber \\
& &\mu_{gs}=[\mu-\frac{1}{2}(-g
A^0_3)+\frac{1}{3}(-\frac{\sqrt{3}}{2}g A^0_8)]-\frac{1}{2}\mu'_3
+\frac{1}{3}\mu'_8 ; \label{muI}
\end{eqnarray}
\begin{eqnarray}
& &\mu_{rs}=[\mu +\frac{1}{2}(-g A^0_3)
+\frac{1}{3}(-\frac{\sqrt{3}}{2}g
A^0_8)] +\frac{1}{2}\mu'_3 +\frac{1}{3}\mu'_8 , \nonumber \\
& &\mu_{bu}=[\mu -\frac{2}{3}(-\frac{\sqrt{3}}{2}g A^0_8)]
-\frac{2}{3}\mu'_8 ; \label{muII}
\end{eqnarray}
\begin{eqnarray}
& &\mu_{gu}=[\mu -\frac{1}{2}(-g A^0_3)
+\frac{1}{3}(-\frac{\sqrt{3}}{2}g
A^0_8)] -\frac{1}{2}\mu'_3 +\frac{1}{3}\mu'_8 , \nonumber \\
& &\mu_{rd}=[\mu +\frac{1}{2}(-g
A^0_3)+\frac{1}{3}(-\frac{\sqrt{3}}{2}g A^0_8)] +\frac{1}{2}\mu'_3
+\frac{1}{3}\mu'_8 ; \label{muIII}
\end{eqnarray}
\begin{eqnarray}
& &\mu_{ru}=[\mu +\frac{1}{2}(-g A^0_3)
+\frac{1}{3}(-\frac{\sqrt{3}}{2}g A^0_8)] +\frac{1}{2}\mu'_3
+\frac{1}{3}\mu'_8 , \nonumber \\
& &\mu_{gd}=[\mu-\frac{1}{2}(-g
A^0_3)+\frac{1}{3}(-\frac{\sqrt{3}}{2}g
A^0_8)]-\frac{1}{2}\mu'_3 +\frac{1}{3}\mu'_8,\nonumber \\
& &\mu_{bs}=[\mu-\frac{2}{3}(-\frac{\sqrt{3}}{2}g
A^0_8)]-\frac{2}{3}\mu'_8 ; \label{muMIX}
\end{eqnarray}
which are essentially equivalent to the conventional expressions of
$\mu_i$. In the CFL phase, $m_s$ takes effect on the Fermion
accumulation in the way of a mismatch between the strange- and
light-flavor Fermion momenta ( see Eq.(\ref{pfcommon}) ). In the
$K^0$ condensed environment, nevertheless, $m_s^2/2\mu$ has been
regarded as the strangeness chemical potential to trigger the
condensation. Thus, the $m_s$ effect on the CFL$K^0$ Fermion
accumulation is expected to become modified in somewhat way.

To treat the $m_s$ effects self-consistently, we examine the
influence of $K^0$ condensate on the quark properties firstly. It is
well known that Goldstone boson condensation(s) may be realized via
axial flavor transformation of quark field
\begin{eqnarray}
q \rightarrow q'=\exp[\frac{i}{2}(\theta_\alpha
\lambda_\alpha)\gamma_5] q.\label{qtrans}
\end{eqnarray}
For our purpose, the indices $\alpha=6,7$ are considered in
Eq.(\ref{qtrans}) while $\sqrt{\theta_6^2 + \theta_7^2}$ is just the
condensate strength angle $\theta$ \cite{buba}. Under this
transformation, the quark-antiquark condensates for three flavors
become
\begin{eqnarray} \langle \overline{u'}u'\rangle = \langle
\overline{u}u\rangle, \,\,\,\, \langle \overline{d'}d'\rangle =
\langle \overline{d}d\rangle \cos\theta, \,\,\,\, \langle
\overline{s'}s'\rangle = \langle \overline{s}s\rangle \cos\theta.
 \label{qqtrans}
\end{eqnarray}
Within the model-independent framework, Eq.(\ref{qqtrans}) may be
attributed  to the modifications in quark masses
\begin{eqnarray}
m'_{u} = m_u, \,\,\,\, m'_{d} = m_d \cos\theta, \,\,\,\, m'_{s} =
m_s \cos\theta,
 \label{mqtrans}
\end{eqnarray}
at the mean-field level.\footnotemark[2] \footnotetext[2] { When the
quark-antiquark condensates ( as well as the diquark condensates )
are taken into account explicitly, however, Eq.(\ref{mqtrans}) no
longer holds valid and thus the method developed in this work is not
available yet. Actually, the descriptions of color superconducting
phases ( including the CFL$K^0$ phase ) have been studied in
NJL-type models ( see, e. g. Refs.\cite{buba,bubsho,bub02} ). There,
the color neutrality is imposed by hand, e. g. by the numerical
tuned values of color chemical potentials, which is very different
from the treatments within the model-independent framework.}
Although themselves are not the quark masses, $m'_{u,d,s}$ do
reflect the actual contributions of quark masses on the Fermi
surface phenomenon in the $K^0$ condensed environment. Ignoring the
light masses, it is not ${m_s}^2/2\mu$  but ${m'_s}^2/2\mu$ to cause
the CFL$K^0$ Fermi-momentum mismatches at the leading order. In
analogy with Ref.\cite{alfj02}, the common Fermi momenta are
rewritten as
\begin{eqnarray} & &p_{F,(gs,bd)}^{com}=[\mu-\frac{1}{4}(-g A^0_3)-\frac{1}{6}(-\frac{\sqrt{3}}{2}g
A^0_8)]-\frac{1}{4}\mu'_3 -\frac{1}{6}\mu'_8
-\frac{{m'_s}^2}{4 \mu}, \nonumber \\
& &p_{F,(rs,bu)}^{com}=[\mu+\frac{1}{4}(-g
A^0_3)-\frac{1}{6}(-\frac{\sqrt{3}}{2}g A^0_8)] +\frac{1}{4}\mu'_3
-\frac{1}{6}\mu'_8
-\frac{{m'_s}^2}{4\mu}, \nonumber \\
& &p_{F,(rd,gu)}^{com}=[\mu+\frac{1}{3}(-\frac{\sqrt{3}}{2}g A^0_8)]
+\frac{1}{3}\mu'_8, \nonumber \\
& &p_{F,(ru,gd,bs)}^{com}=\mu-\frac{{m'_s}^2}{6\mu} ; \label{pfcom2}
\end{eqnarray}
where Eqs.(\ref{muI}-\ref{muMIX}) have been considered.

With the help of $\mu_i$ and $p_{F,i}^{com}$, the free energy
relevant for the CFL$K^0$ Fermion accumulation reads
\begin{eqnarray}
\Omega'_{CFL}(\mu'_3,\mu'_8,m'_s)=\frac{1}{\pi^2}\sum
\int_0^{p_{F,i}^{com}} (\sqrt{p^2+{m'_s}^2}-\mu_i) p^2 d p
 + \frac{1}{\pi^2}\sum \int_0^{p_{F,i}^{com}} (p-\mu_i) p^2 d p,
 \label{Ocfl'}
\end{eqnarray}
where the first term of RHS involves the sum of $i=gs,rs,bs$ while
the second term does the sum of $i=bd,bu,rd,ru,gu,gd$. Although
Eq.(\ref{Ocfl'}) has the similar form as the CFL free energy,
$m'_s$, $\mu'_{3}$ and $\mu'_{8}$ have replaced $m_s$, $\mu_{3}$ and
$\mu_{8}$ to become the variables respectively while
 $A^0_{3}$ and $A^0_{8}$ are actually the invariables being
independent of $\mu'_{3,8}$. As in Ref.\cite{alfj02}, we concern the
$\mu'_{3,8}$-related terms up to order $m_s^4$ to the purpose of
yielding the leading-order solutions. Explicitly, the
strange-quark-involved part in Eq.(\ref{Ocfl'}) reads
\begin{eqnarray}
-\frac{1}{12\pi^2}\sum [{p_{F,i}^{com}} {\mu_i^3} -
\frac{5}{2}{p_{F,i}^{com}}{\mu_i}{{m'_s}^2}] +
\frac{3}{8\pi^2}{{m'_s}^4}\log(\frac{m'_s}{2\mu})+\ldots ,
 \label{Ocfl'1}
\end{eqnarray}
while the light-quark-involved part is
\begin{eqnarray}
-\frac{1}{12\pi^2}\sum {p_{F,i}^{com}} {\mu_i^3}.
 \label{Ocfl'2}
\end{eqnarray}
Further expanding the above equations, it is found that there are
 the components like $A^0_{3,8}{(\mu'_{3,8})}^2\mu$, ${(A^0_{3,8})}^2
\mu'_{3,8} \mu$ and $A^0_{3,8}{\mu'_{3,8}} {{m'_s}^2}$ in our
concerned terms. Noticing that $\mu'_{3,8}$ and $A^0_{3,8}$ have
order $m_s^2$, the order of these components is beyond $m_s^4$ and
they are actually irrelevant for evaluating $\mu'_{3,8}$. To the
order at which we are working, it is practical to ignore the
$A^0_{3,8}$-related terms in the expansion of $\Omega'_{CFL}$.
Equivalently, we can simplify the square-bracket parts in the
expressions of $\mu_i$ and $p_F^{com}$ ( i. e.
Eqs.(\ref{muI}-\ref{muMIX}) and (\ref{pfcom2}) ) as $\mu$. Under the
above approximation, it becomes easy to yield the
Alford-Ragagopal-type solutions $\mu'_{3,8}$. If replacing $\mu_3$,
$\mu_8$ and $m_s$ by $\mu'_3$, $\mu'_8$ and $m'_s$ respectively, it
is obvious that the conventional CFL expressions of $\mu_i$ and $p
_{F,i}^{com}$ hold ( approximately ) unchanged in the CFL$K^0$ case.
Correspondingly, $\mu'_{3}$ and $\mu'_{8}$ should have the formally
same result as Eq.(\ref{ncfl}) as long as the replacements $\mu_3
\rightarrow \mu'_3$, $\mu_8 \rightarrow \mu'_8$ and $m_s \rightarrow
m'_s$ are considered. Instead of evaluating
${\partial\Omega'_{CFL}}/{\partial \mu'_{3,8}}=0$, we find that the
fictional chemical potentials are
\begin{eqnarray} \mu'_3=0, \,\,\,\,\,\, \mu'_8=-\frac{{m'_s}^2}{2\mu},
\label{ncflk}
\end{eqnarray}
at the leading order, by comparing with the well-known result
Eq.(\ref{ncfl}).

As an example, we examine the case of the maximal kaon condensate (
$\theta=\pi/2$ ). In such a CFL$K^0$ phase, the value of $m'_s$
becomes zero so that its effect on the common Fermi momenta
vanishes. Thus, the introduction of $\mu'_{3,8}$ is no longer
necessary and we have $\mu'_3=\mu'_8=0$, which is consistent with
Eq.(\ref{ncflk}). Correspondingly, the factual color chemical
potentials become $\mu_3 =- g A^0_3 = -m_s^2/4\mu$ and $\mu_8=
-\frac{\sqrt{3}}{2}g A^0_8= -m_s^2/8\mu$, i.e. Eq.(\ref{link}) is
reasonable. Even if so, it is interesting to investigate the Fermion
surface phenomenon in the resulting color-neutral phase. From
Eq.(\ref{pfcom2}), the common Fermi momenta are found to become
$p_{F,(gs,bd)}^{com}=\mu+ {m_s^2}/{12\mu}$,
$p_{F,(rs,bu)}^{com}=\mu- {m_s^2}/{24\mu}$,
$p_{F,(rd,gu)}^{com}=\mu- {m_s^2}/{24\mu}$, and
$p_{F,(ru,gd,bs)}^{com}=\mu$. Therefore, the average Fermi momentum
for all the paired quarks is equal to $\mu$ in the maximal $K^0$
condensed phase ( if simply adopting our approximation, indeed, it
is obvious that the common Fermi momenta are reduced to $\mu$ ).
This result is different from the conventional case where the
average Fermi momentum is $\mu-m_s^2/6\mu$, but comparable with the
ideal CFL case in the absence of $m_s$. In this sense, we conclude
that the Fermion accumulation in the maximal $K^0$ condensed
environment behaves like that in the ideal CFL matter. Physically,
the reason lies in the fact that the strange quark mass does not
involve the Fermi surface phenomenon directly since its effect
causes the maximal kaon condensation completely. This conclusion was
not drawn in the previous literatures yet. Actually, one would have
$p_{F,(gs,bd)}^{com}=\mu - {m_s^2}/{6\mu}$,
$p_{F,(rs,bu)}^{com}=\mu- {7m_s^2}/{24\mu}$,
$p_{F,(rd,gu)}^{com}=\mu- {m_s^2}/{24\mu}$ and
$p_{F,(ru,gd,bs)}^{com}=\mu- {m_s^2}/{6\mu}$ by inserting the color
neutral solutions into Eq.(\ref{pfcommon}). There, the $m_s$ effect
is improper to be counted twice and the resulting Fermion
accumulation is not correct. Thus, the treatment developed in the
present work is necessary to avoid the double counting on the $m_s$
effect and illuminate the $K^0$ condensed phase self-consistently.

In principle, there are possibilities except for the maximal $K^0$
condensate. In realistic situations, e.g. for not-very-large $\mu$,
the instanton contribution on the kaon mass needs to be taken into
account \cite{ins}. In this case, the $K^0$ condensation is
suppressed ( and even no longer occurs ) in CFL quark matter. The
$K^0$ condensation ( if it exists ) has an arbitrary strength
$\theta$ even if in the situation of $m_s>> m_{u,d}$. In the
CFL$K^0$ phase, the nonzero $m'_s$ leads to the nonvanishing
$\mu'_8$ ( see Eq.(\ref{ncflk}) ). Therefore, Eq.(\ref{link}) no
longer holds valid and the factual color chemical potentials become
\begin{eqnarray}
& & \mu_3=-\frac{m_s^2}{4\mu}+ \frac{m_s^2}{4\mu}\cos\theta, \nonumber
\\
& & \mu_8= -\frac{m_s^2}{8\mu}- \frac{3m_s^2}{8\mu}\cos\theta -
\frac{m_s^2}{2\mu}{(\cos\theta)}^2,  \label{ncflkk}
\end{eqnarray}
at the leading order. In the resulting color-neutral phase the
average Fermi momentum becomes $\mu-{m'_s}^2/6\mu$, which manifests
that the Fermi surface behavior is influenced by $m_s$ partly.
Although it was seldom discussed in the literature, such a CFL$K^0$
phase is likely to exist in realistic situations. More importantly,
it is definitely pointed out that the Kryjevski 's ( expectation
values of ) gluon fields do not correspond to the factual color
chemical potentials. In fact, the Fermi surface phenomenon
influenced by kaon condensation leads to the extra color static
potentials, which are independent from $A^0_{3,8}$ and connected
with the $K^0$ condensate strength. These conclusions clarify some
presumptions on the $K^0$ CFL condensed phase in the literature and
might be important for fully understanding the unconventional CFL
phases. The present method could be extrapolated if more physics
involving Goldstone mode condensations is considered. When taking
the electron chemical potential and the light quark masses into
account, the electric-charged boson condensations and electric
neutrality of the CFL matter need to be examined. As stressed in
this work, both the dynamics of the condensed mode and the
Fermi-surface behavior of color-flavor locked quarks should take
effect on the color/electric neutrality.

In summary, we investigate color neutral problem in the CFL quark
matter with arbitrary $K^0$ condensate. In order to treat the $m_s$
effect self-consistently, we introduce the fictional variables
$\mu'_{3,8}$ and $m'_s$ and reexamine the Fermi-surface behavior in
$K^0$ condensed environment. In this way, the calculations of
$A^0_{3,8}$ ( from the Goldstone-mode Effective Lagrangian ) and
$\mu'_{3,8}$ ( from the Fermion free energy $\Omega_{CFL'}$ ) become
independent from each other. The factual color chemical potentials
are obtained, which implies the breaking of the previous ansatz
Eq.(\ref{link}). Correspondingly, the Fermion accumulation in $K^0$
condensed environment is investigated and it is found to differ from
the conventional case. Since the results of $A^0_{3,8}$ are adopted
directly, our treatment  is actually based on the Alford-Ragagopal
method and is valid only at the mean-field level. It is still
unclear how to understand the present solutions ( in particular the
extra chemical potentials $\mu'_{3,8}$ ) in the framework of the "
full " effective field theory including both Goldstone bosons and
the paired quarks \cite{kry}. Furthermore, the intrinsic link
between our discussed CFL$K^0$ phase and the $p-$wave $K^0$
condensed phase \cite{sch06} needs to examined seriously. Another
important issue that not discussed in this work involves the gapless
formation. In the presence of $m_s$, the gapless CFL phase was
predicted in Ref.\cite{alf04} and the influence of maximal $K^0$
condensate on it was investigated in Ref.\cite{zhang,kry}. With the
color neutral solutions Eq.(\ref{ncflkk}), the gapless formation
might warrant further investigation in the environment with
arbitrary $K^0$ condensate, which is beyond the scope of the present
work.

\vspace{1.0cm} \noindent {\bf Acknowledgements} \vspace{0.5cm}

This work was supported by National Natural Science Foundation of
China ( NSFC ) under Contract No.10405012.

\end{document}